\documentstyle[11pt,newpasp,twoside,epsf]{article}
\begin{document}
\title{The Mass Function of the Arches Cluster from Gemini Adaptive Optics Data}
 \author{Andrea Stolte}
\affil{Max-Planck-Institute for Astronomy, Heidelberg, Germany \hspace{3cm} 
European Southern Observatory, Garching, Germany}
 \author{Eva K. Grebel}
\affil{Max-Planck-Institute for Astronomy, Heidelberg, Germany}
 \author{Wolfgang Brandner}
\affil{European Southern Observatory, Garching, Germany}
 \author{Don F. Figer}
\affil{Space Telscope Science Institute, Baltimore, USA}

\begin{abstract}
We present the mass function (MF) of the Arches cluster obtained from 
ground-based adaptive optics data in comparison with results derived 
from HST/NICMOS data. A MF slope of $\Gamma = -0.8 \pm 0.2$ is obtained.
Both datasets reveal a strong radial variation 
in the MF, with a flat slope in the cluster center,
which increases with increasing radius.
\end{abstract}

\section{Introduction}
The Arches cluster is one of the most massive young star clusters found in 
the Milky Way. The total stellar mass is estimated to be $\ga 10^4 M_\odot$, 
with a central stellar density of $3 \cdot 10^5 M_\odot\ {\rm pc}^{-3}$ 
(Figer et al. 1999).
Located at a projected distance of only 25 pc from the Galactic
Center (GC), it evolves under extreme physical conditions.
As the formation of massive stars ($M > 30 M_\odot$) 
is proposed to be strongly dependent on high stellar and gas densities
(Behrend \& Maeder 2001, Bonnell et al. 1998),
star clusters close to the GC are important test beds for star formation (SF) 
scenarios in dense regions.
Displaying the highest stellar density found in a massive young cluster (YC) 
in the Milky Way, the Arches cluster is a unique object for the comparison with 
massive SF models. 
Due to the proximity to the GC, strong tidal forces act towards the 
disruption of the cluster entity, causing the timescale for dynamical 
evolution to be short, which may lead to the dissolution of the cluster
within $10-20\,{\rm Myr}$ (Kim et al. 1999, Portegies Zwart et al. 2002). 
With an age of only $2.5 \pm 0.5$ Myr (Figer et al. 2002, in prep.),
the stellar population of the Arches cluster 
should be mostly unaffected by stellar evolution,
but the spatial appearance of the cluster should reflect the fast dynamical 
evolution.

\section{Observations}
We analyse high-resolution adaptive optics (AO) $H$ and $K_s$-band 
observations of the Arches cluster center obtained 
during the Gemini North science verification\footnote{The Gemini
North science verification data are publicly available at http://www.gemini.edu} 
(SV) with the University of Hawai'i AO system Hokupa'a (Graves et al. 2000). 
These data are compared to HST/NICMOS observations
of the same field obtained by Figer et al. (1999).
In uncrowded regions, the spatial resolution and limiting magnitudes
in both datasets are comparable given a 3 to 4 times longer 
integration time in $H$ and $K_s$, respectively, for the Gemini observations.
In crowded regions, the AO data are limited by uncompensated seeing haloes
from bright stars caused by the low Strehl ratio of only a few percent 
in the Gemini SV data. These haloes locally decrease the 
detection efficiency of faint sources. Evidence for this is seen at the faint end 
of the colour-magnitude diagrams and in the MF of the dense cluster center. 
\begin{figure}
\plotfiddle{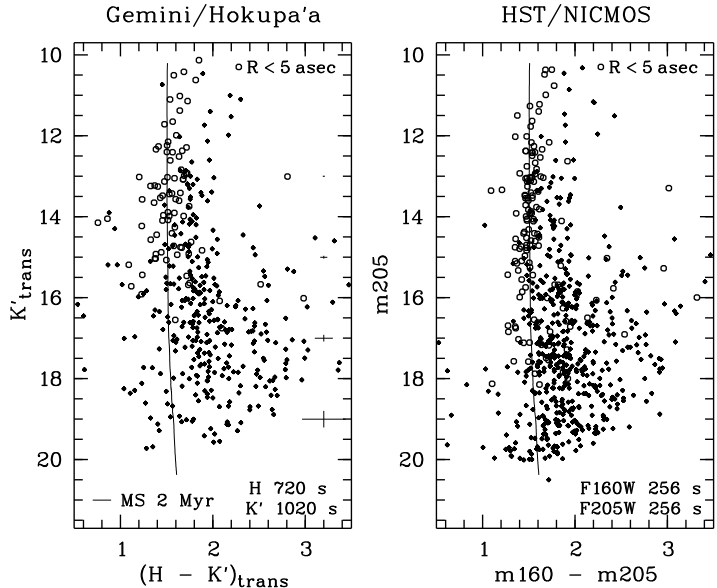}{7cm}{0}{55}{55}{-160}{-220}
\caption{Arches colour-magnitude diagrams obtained with the Gemini North
AO system Hokupa'a and HST/NICMOS. Stars in the cluster center are marked
as circles. The 2 Myr isochrone (Lejeune \& Schaerer 2001) used to transform
magnitudes into stellar masses is shown as a solid line.}
\end{figure}

\section{Colour-Magnitude Diagrams}
The $H-K, K$ colour-magnitude diagrams (CMDs) are shown in Fig. 1.
The dominant feature in the CMDs is the Arches main sequence (MS). 
Due to the low Strehl ratio in the Gemini data, the upper part of the MS, 
corresponding to the bright
population in the cluster center, appears less confined than in the 
NICMOS CMD. The fainter stars, not detected in the cluster center due to crowding, 
show a large scatter towards redder colours, which is caused by a strong 
colour gradient observed over the Arches field.
\begin{figure}
\plotfiddle{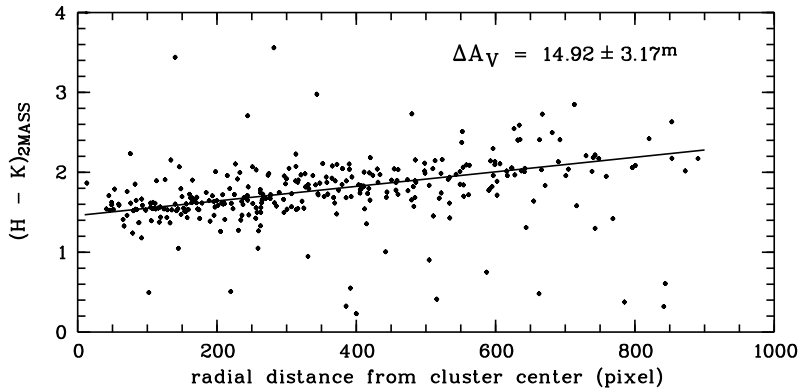}{5cm}{0}{65}{65}{-214}{-360}
\caption{Colour gradient observed over the Gemini field of view of 20\arcsec (1024 pixels). The change in $H-K_s$ is shown as a function of distance from the cluster center.}
\end{figure}

\section{Results}

\subsection{Colour Gradient}
The colour trend is apparent in both datasets, with increasingly
red stellar colours observed at increasing radii (Fig. 2).
The change in $H-K_s$ corresponds to
an increase in visual extinction of $\Delta A_V \sim 15$ mag
($\Delta A_V = \Delta (A_H-A_K)/0.0630$, Rieke \& Lebofsky (1985)).
The well confined main sequence
at the bright end of the CMD shows that the innermost  
cluster center ($\sim 5\arcsec$, 0.2 pc at 8 kpc distance) 
is unaffected by this colour trend. This suggests that the remaining 
dust and gas has been expelled from the cluster center by winds
and/or photoevaporation from the massive stars.

\subsection{Mass Function}
After correction for the extinction variation, 
a 2 Myr isochrone from the Geneva basic set of stellar evolution models
(Lejeune \& Schaerer 2001) has been used to transform $K$-band 
magnitudes into stellar masses. The resulting MF (Fig. 3) displays a 
slope of $\Gamma = -0.8 \pm 0.2$ in the mass range $10 < M < 65 M_\odot$ 
in the Gemini as well as the HST dataset. This slope is
flatter than the Salpeter slope of $\Gamma = -1.35$, often assumed to be 
universal in young star forming regions.
When neglecting the colour trend, we derive a MF slope 
as flat as $\Gamma \sim -0.5 \pm 0.2$. This clearly shows that the 
changes in extinction in SF regions are not negligible
when deriving mass functions. The flat MF observed in 
Arches indicates an overdensity of high-mass stars as compared 
to other SF regions, thereby supporting models predicting
enhanced high-mass SF in dense regions. 

We have analysed the stellar population at different radii from the 
cluster center to derive the radial variation in the MF (Fig. 4).
The MF slope appears very flat in the
cluster center, where stellar and gas densities are highest,
and increases with increasing radius. While mainly
massive stars are found in the central part of Arches, an increasing 
number of faint, low-mass stars is found in the cluster outskirts,
indicating mass segregation. 
To exclude that the derived variation is due to crowding,
incompleteness simulations have been performed in each radial bin, 
shown as dotted lines in Figs. 3 and 4. Although incompleteness 
begins to affect the MF earlier in the cluster center ($R < 5\arcsec$),
the corrected MF remains very flat for $M > 10 M_\odot$.
From the distribution of stellar masses, 
we can obtain a rough estimate of the relaxation timescale,
$t_{rh} = \frac{6.63 \cdot 10^8 {\rm yr}}{\ln(0.4N)}\left(\frac{M}{10^5\,M_\odot}\right)^{1/2}\left(\frac{1\,M_\odot}{m_\ast}\right)\left(\frac{r_hm}{1\,{\rm pc}}\right)^{3/2}$,
where $M$ is the mass within the half-mass radius $r_{hm}$, 
$m_\ast$ the characteristic mass, and N the number of stars
(Binney \& Tremain 1987). We obtain $r_{hm} = 0.4\,{\rm pc}$ and 
observe $M(r_{hm}) = 6300\,M_\odot$ as a lower limit
to the total mass. The median mass within $r_{hm}$
is measured to be $7.3\,M_\odot$.
Assuming a total mass of at least $10^4\,M_\odot$ leads to 
$t_{rh} = 1.2\,{\rm Myr}$. A characteristic mass of $1\,M_\odot$, more typical
for SF regions, yields $t_{rh} = 6.2\,{\rm Myr}$. With $t_{rh}$
being on the same order of magnitude as the age of Arches, 
it is impossible to discrimate between primordial and dynamical 
mass segregation.
\begin{figure}
\plotfiddle{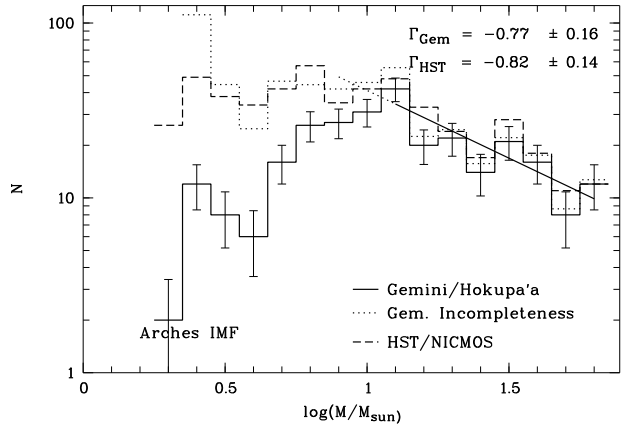}{5cm}{0}{90}{90}{-140}{-480}
\caption{Mass function of the Arches cluster. $\Gamma$ denotes the 
slope of a linear, weighted least-squares fit to the bright end of the MF,
Salpeter would be $\Gamma = -1.35$.}
\end{figure}
\begin{figure}
\plotfiddle{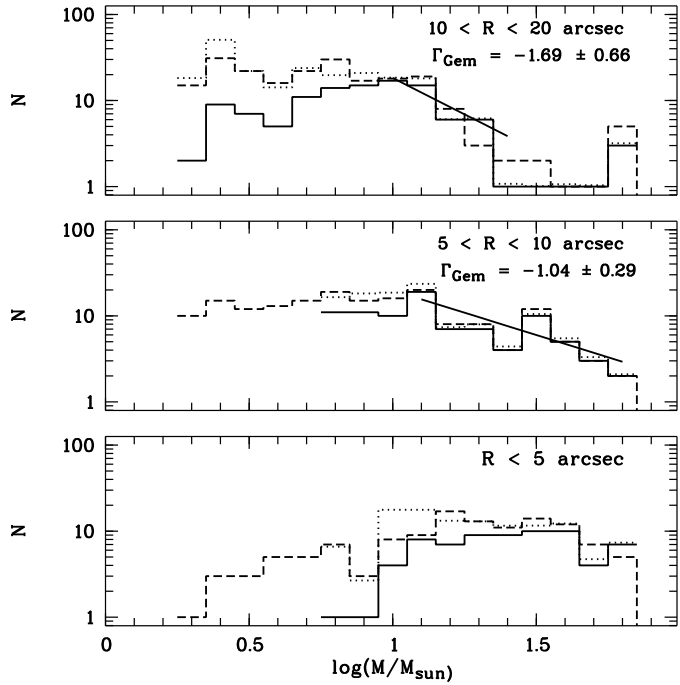}{8.5cm}{0}{50}{50}{-160}{-30}
\caption{Radial variation of the Arches mass function. Symboles as in Fig. 3.}
\end{figure}

\section{Summary}
The radial variation in the MF of the Arches cluster as well as 
the flat integrated mass function support massive cluster and star
formation models, suggesting high-mass stars to form in dense
cluster environments. Although the radial variation in the MF
strongly suggests mass segregation in Arches to be present,
the short dynamical timescales, strongly influenced by the GC tidal field,
prohibit discrimination between primordial and dynamical segregation effects.

The close similarity of the mass functions obtained from the Gemini
AO and the HST datasets reveal that high-resolution ground-based
AO data are capable to produce comparable physical results as 
space-based observations - in particular in dense regions where 
spatial resolution is essential. 

\acknowledgements
Based on observations obtained with the Gemini North telescope and HST.

\end{document}